\documentclass[USenglish,twocolumn]{article}

\usepackage[utf8]{inputenc}
\usepackage{graphicx}

\begin{document}




  \title{ Relativity, Anomalies and  Objectivity Loophole in Recent Tests of Local Realism\\
Adam Bednorz\footnotemark[1]}








\maketitle
\footnotetext[1]{Faculty of Physics, University of Warsaw, ul. Pasteura 5, PL02-093 Warsaw, Poland, E-mail: Adam.Bednorz@fuw.edu.pl}

 \begin{abstract}
{Local realism is in conflict with special quantum Bell-type models. Recently, several experiments have demonstrated violation of local realism 
if we trust their setup assuming special relativity valid. In this paper we question the assumption of relativity, point out not commented anomalies and show that the experiments have not closed objectivity loophole because clonability of the result has not been demonstrated. We propose several improvements in further experimental tests of local realism make the violation more convincing.}
\end{abstract}

\section{Introduction}

In contract to classical physics, quantum mechanics does not provide a direct realistic interpretation.
In realism, there exists a joint probability for all possible measurements, even if not simultaneously feasible.
The measurements can be choice-dependent. If the choice is located at some spacetime point then local realism means
that the probability cannot depend on remote choices -- no signaling. If we stick to relativity, remote means spacelike, i.e. beyond the reach of light.
Otherwise, remote means simply lack of possible communication (due to either sufficiently short time, large enough distance or lack of known physical communication channel).

Local realism has been questioned by  Einstein, Podolsky and Rosen (EPR) \cite{epr}, later put by Bell \cite{bell} in the form of special inequality, involving at least two
observers each choosing between two measurements \cite{chsh}. Local realism defined by Bell requires existence of a the joint probability distribution for all choice-dependent outcomes. Nowadays, there are many other Bell-type tests \cite{bell,eber}. Bell inequality is violated in a simple quantum model.
However, it is not so simple to implement the model experimentally. One has to face a lot of problems, such as low detection efficiency,
decoherence and fast choice and readouts. Only recently, the experiments started to overcome these problems \cite{hensen,nist,vien,munch}. Nevertheless,
the raw data from these experiments shed light on further possible problems, questioning both the interpretation of the results and completeness of the underlying theory \cite{ab17,khlar}.

In this paper we summarize these experiments and the current status of local realism and relativity. We will explain how to improve next experiments
to check violation of no-signaling and relativity. In addition we will show the so-called \emph{objectivity loophole}. Objectivity means that the result of the experiment must get ensured to be objective -- clonable arbitrary number of times \cite{broad}. Otherwise, one imagine that Eve, a hacker or some uncontrolled physical process alters insecure data to violate a Bell-type inequality. In this case the time tag of the readout should be shifted to the end of the supposed hack which may be enough late to get informed  about other party's choice within lightcone. Such hacking is less likely when several copies exist. Recent experiments have not attempted even making two copies of the results. This is complementary to the freedom of choice loophole. The latter occurs at the start of the experiment and we can only narrow it by complicated random number generations, but never closed absolutely due to conspiracy in the past. The objectivity loophole occurs at the end of experiment and can be only narrowed by sufficiently large number of independent, secure copies, but never closed completely for an extraordinarily skilled hacker. In fact the loophole is relevant not only for EPR or Bell experiments but also all other experiments, where we want to prove that the readout has been
fixed at certain time.

The paper is organized as follows. We start from recalling Bell-type tests used in recent experiments and their assumptions.
Then we discuss no-signaling principle in context of relativity and the experimental data. Next we explain insufficient modeling of the recent tests
and possible technical issues. Finally, we describe the objectivity loophole and suggestions how to close or narrow it.

\begin{figure}
\includegraphics[scale=.5]{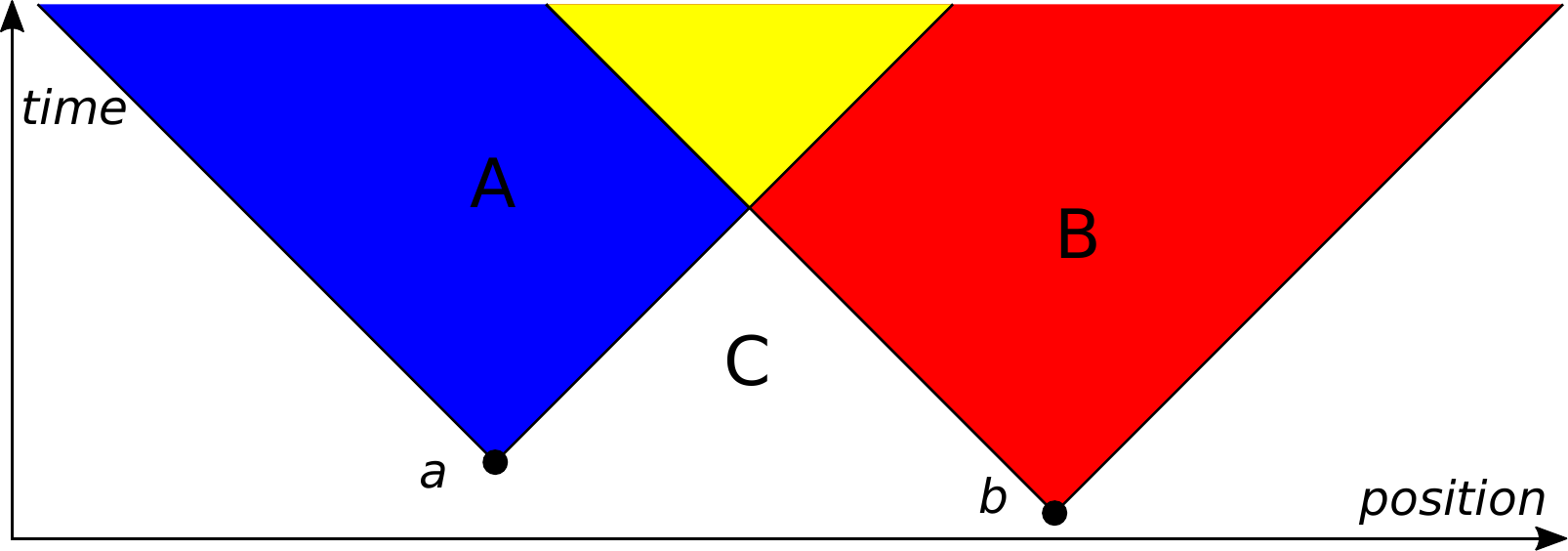}
\caption{Spacetime picture of a simple Bell-type test of local realism. Free choices, denoted by small letters generate spatio-temporal lightcones, bounded by speed of light. The measurements, denoted by capital letters must be completed outside of the other party's choice ($A$ and $B$) or both ($C$).}
\label{spt}
\end{figure}

\section{Bell-type tests}

The most known Bell test is due to Clauser, Horne, Shimony, Holt (CHSH) \cite{chsh}, for two observers $A$ and $B$.
Each of them can choose the measurement between $1$ and $2$ ($A$) or $3$ and $4$ ($B$) the choice-dependent outcome
$A_i$ and $B_j$ has values $\pm 1$. Locality means that outcomes depend only on local choice, i.e. $A_{13}=A_{14}=A_1$, etc.
This assumption is critical because sufficiently slow choice or readout makes it invalid -- the observers can ultimately get informed about the other party's choice. One usually refers here to relativistic bound of speed of light, see Fig. \ref{spt} but the examples below do not at all rely on relativity. 
They are just simple quantum models with assumed states and their evolution.

Then realism requires existence of the joint positive probability $p(A_1,A_2,B_3,B_4)$, yielding the inequality
\begin{equation}
-2\leq \langle A_1B_3\rangle+\langle A_1B_4\rangle+\langle A_2B_3\rangle-\langle A_2B_4\rangle\leq 2\label{chshin}
\end{equation}
It follows from the fact that $A_1(B_3+B_4)+A_2(B_3-B_4)$ is always $\pm 2$. Quantum rule, $\langle AB\cdots Z\rangle=\mathrm{Tr}\hat{A}\hat{B}\cdots\hat{Z}\hat\rho$
for the simultaneously measurable observables (Hermitian) $\hat{A}$, $\hat{B}$ etc., and the (positive, Hermitian, and normalized) state matrix $\hat\rho$, allows
to predict all here relevant correlations.
Taking $\hat{\rho}=|\psi\rangle\langle\psi|$ with  Bell states $\sqrt{2}|\psi\rangle=|+-\rangle-|-+\rangle$ in the basis $|\pm_A,\pm_B\rangle$ and the observables $\hat{X}_i=e^{i\phi_i}|+\rangle\langle -|+e^{-i\phi_i}|-\rangle\langle +|$ (in the respective $X=A,B$ subspace) we get $X_i=\pm 1$
and $\langle A_iB_j\rangle=-\cos(\phi_i-\phi_j)$. For $\phi_{1,2,3,4}=(0,\pi/2,5\pi/4,3\pi/4)$, the inequality (\ref{chshin}) is violated as the middle correlator reads
$2\sqrt{2}\simeq 2.84>2$.
Unfortunately, experimental implementation of this model is quite hard. Either the actual outcome has the third value $0$ (usually meaning lost particle)
or the state decoheres too quickly. In both cases the original inequality is in practice no longer violated. There are at least two
counteractions. One is to condition the state by entanglement swapping \cite{swap}. The distinguish between main Bell (localized) uppercase states $|\pm_{A,B}\rangle$ and auxiliary communication (traveling) lowercase states $|\pm_{a,b}\rangle$ If observers first generate their local Bell states
$\sqrt{2}|\psi_X\rangle=|+_X,-_x\rangle-|-_X,+_x\rangle$  for $X=A,B$, $x=a,b$ (matching upper and lower case letter) and send the lower case $a,b$ states to the observable
$\hat{C}=|\phi\rangle\langle\phi|$ with $\sqrt{2}|\phi\rangle=|+_a,-_b\rangle-|-_a,+_b\rangle$ then the outcome $C=1$ heralds the Bell state between $A$ and $B$.
Then even very low efficiency of the heralding does not prevent the Bell test if $C$ is local - independent of all choices, i.e. $C=C_{13}=C_{14}=C_{23}=C_{24}$. The procedure is preselection and not postselection and so
violation of $C=1$ conditioned (\ref{chshin}) is still correct signature of violation of local realism.

Another solution is to replace CHSH inequality with Eberhard inequality \cite{eber} which takes better into account detection efficiency ($0$ outcomes).
The setup is similar to the previous one only $X_i=0,1$ (the case $-1$ is reassigned also to $0$). Positive $p$ yields the inequality
\begin{equation}
p(1_1,1_3)-p(1_1,0_4)-p(0_2,1_3)-p(1_2,1_4)\leq 0\label{eber}
\end{equation}
It follows from the decomposition of probability $p(1_1,1_3)=p(1_1,1_2,1_3,0_4)+p(1_1,0_2,1_3)+p(1_1,1_2,1_3,1_4)$
while $p(1_1,0_4)\geq p(1_1,1_2,1_3,0_4)$, $p(0_2,1_3)\geq p(1_1,0_2,1_3)$ and $p(1_2,1_4)\geq p(1_1,1_2,1_3,1_4)$.
Applied to the ideal Bell state with $-1$ replaced by $0$ we get $p(1_i,1_j)=(1-\cos(\phi_i-\phi_j))/4$ and
$p(1_i,0_j)=(1+\cos(\phi_i-\phi_j))/4$ and the inequality is violated for the same set of angles with the left hand side reading
$1/\sqrt{2}-1/2\simeq 0.2>0$. 

\section{No-signaling and relativity}

The local realism assumption of readout independence of remote choice is usually based on relativity. 
Yet without relativity, there is a universal causality principle that the readout can depend only on earlier choices.
Combined now with Lorentz invariance, the dependence is restricted to the light cone, namely $r<ct$ where $r/t$ is the spatial distance/time between the choice and readout, $c$ -- speed of light. In relativistic quantum mechanics causality implies compatibility of Heisenberg operators beyond the mutual lightcone (Wightman axiom) \cite{wight}. It also leads to no-signaling principle: probability of one observer readout (not correlation with the other one) cannot depend on remote choices.
In the Bell case it means $p(A_{14})=p(A_{13})$, $p(A_{23})=p(A_{24})$, $p(B_{13})=p(B_{23})$, $p(B_{14})=p(B_{24})$. For entanglement swapping we have 
additionally $p(C_{13})=p(C_{14})=p(C_{23})=p(C_{24})$ and even $p(A_{13},C_{13})=p(A_{14},C_{14})$ etc. No-signaling is directly verifiable in the Bell test and serves as a partial check if locality is indeed true (it cannot rule out hidden signaling affecting only correlations but then anyway the claim of violation of local realism is justified).

In the data of recent tests, two experiments show moderate violation of no signaling \cite{ab17}, noted also by others \cite{khlar}. In Delft \cite{hensen} $p(C_{ij})$ is not equal with $95$\% confidence level while $p(1_{13})\neq p(1_{14})$ in NIST \cite{nist} with similar confidence level. In particular, in Delft $N(C_{14}=1)=79$ while $N(C_{24}=1)=51$. Here $N(X_{ij})$ is the number of actually recorded events, while the prediction is $N(X_{ij})=p(X_{ij})N$ for the total number of events $N$. The difference is even larger when removing the special time window for $C=1$ event, then $N(C_{14}=1)=218$ while $N(C_{24}=1)=159$. In NIST, $N(A_{13}=1)=502339$ while $N(A_{23}=505163)$ outside the peak region. In Delft violation would be in conflict with relativity while in NIST
signaling may be also slower than light although there is no visible borderline. Further experiments and more data are necessary to confirm this finding.
It still may be just statistical error.

On the other hand relativity seems to be in conflict with any quantum realism even without Bell example (which -- again -- does not rely on it).
The reason is vanishing of some correlations in vacuum (an invariant state) \cite{ab15,ab16}. For any four-current in spacetime $j(x)$, defining 
$j(p)=\int d^4x e^{ix\cdot p}j(p)$ with Minkowski dot product $a\cdot b=a_0b_0-\vec{a}\cdot\vec{b}$ (fourvectors $a=(a^\mu)=(a_0,\vec{a})$ with spatial vector
$\vec{a}$) invariance implies
\begin{equation}
\langle j^\mu(p)j^\nu(q)\rangle=\delta(p+q)G^{\mu\nu}(p)\label{cor}
\end{equation}
with $G^{\mu\nu}=p^\mu p^\nu\xi+g^{\mu\nu}\eta$ with $\xi$ and $\eta$ depending only on $p\cdot p$. If realism (positive probability
and correlations) holds -- even without quantum mechanics, then $0>\eta>-(p\cdot p)\xi$ for $p\cdot p>0$ (timelike)
and $\eta=0$, $\xi>0$ for $p\cdot p<0$ (spacelike). Imposing additionally charge conservation $j\cdot p=0$ we have then $\xi=0$ so (\ref{cor}) vanishes completely in this case. This is a serious problem for relativistic quantum field theory because once this correlation is zero, any higher order correlation, involving this quantity must also vanish which is impossible to achieve in a natural attempt of quantum realism \cite{ab15,ab16}. In addition,
quantum field theory is often used in a perturbative way, with small interaction coupling. However, no-signaling principle is non-perturbative \cite{wight}.
Any correction violating relativity will keep speed of light as a signaling limit perturbatively while broken nonperturbatively \cite{ab16}.

It must be also stressed that the recent experiments have not tested relativity, only assumed it. This is formally a loophole although relativity is indeed a fair assumption. Nevertheless, the setups contain obvious possible communication channels
(fiber glass) where the information speed is even slower than light using current description. If relativity is wrong, some kind of  hidden (so far)  superluminal interaction can
use this channel. It would be interesting to close these channels during the experiment (single event) to prevent any, even superluminal, interaction using 
them. Then realistic explanation would require nonlocality in the sense of outer communication channels not dedicated to the particular experimental setup.

\section{Anomalies}

\begin{figure}
\includegraphics[scale=.5]{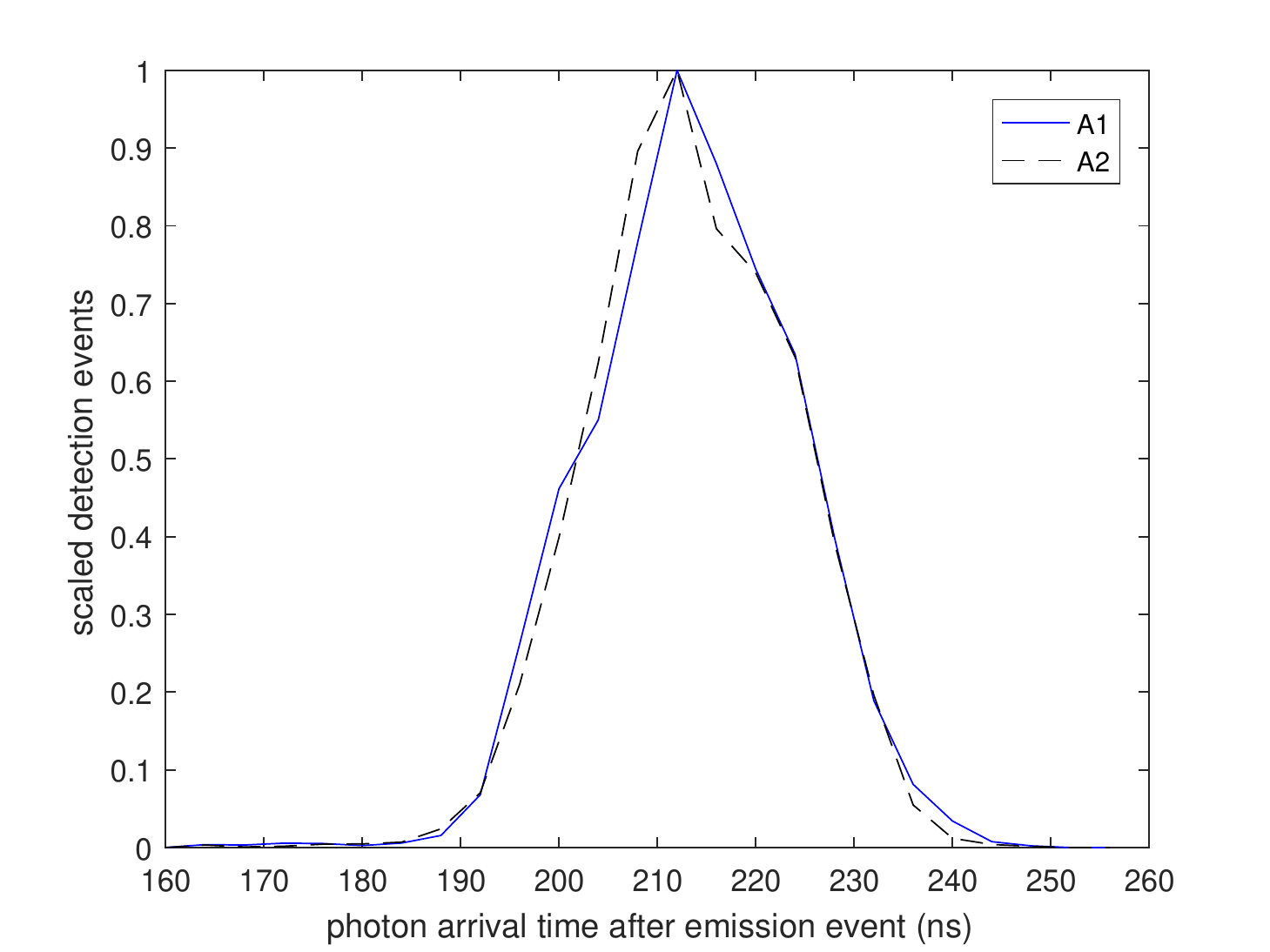}
\caption{Scaled number of detection events at $A$  in Vienna depending on local choice, in $4$ns bins.}
\label{nha}
\end{figure}

In experimental efforts to violate local realism we have to remember that the violation
is based on specific quantum model which should effectively describe the results. Of course local realism can be violated regardless the model behind 
(we compare the data with a classical local realistic inequality) but we also want to do it in a controllable way to be useful in future, e.g. in quantum cryptography. It is therefore relevant to check how accurate is the theoretical quantum model describing the setup and the data. More importantly,
this accuracy boosts the trust in the last, unverifiable, assumption -- freedom of choice. Currently the choice is at least partially made by a quantum random number generator and we have to trust that its choices are not affected by the remote party. One usually models the data by two-dimensional Hilbert space for each of observers giving $4=2\times 2$ total dimension. Then the detector distinguishes only between two states. The probability of a given output 
depends only on the local choice. The ratio between probabilities for different choices, $p(A_1=1)/p(A_2=1)$ cannot depend on any other parameter, also when
binning the time-tags of detection. However, in Vienna \cite{vien} clearly the ratio depends on the binning, see Fig. \ref{nha} (or \cite{ab17} for $B$).The only solution within quantum mechanics is to
\emph{increase} the dimension of the space. It is not particularly unreasonable, e.g. photons can have additional orbital degrees of freedom, apart from two-state polarization, namely their direction of motion. The Pockels cell, normally changing only polarization, may also change the direction and the final photodetector efficiency may depend on this direction. Formally it can be realized by introducing auxiliary states $|k\rangle$, $k=0,...n-1$ for each observer.
then the total space basis is $|j,\pm;k,\pm\rangle$ ($2n\times 2n$ states). Now suppose the initial Bell state reads
$|+0,-0\rangle-|-0,+0\rangle$ and choice makes a local unitary rotation $\hat{U}$ in the basis $\pm$
but also rotation $|j\rangle\langle 0|+|0\rangle]\langle j|$ for $j$ equal the choice index (alternatively, one of choices can lead to null, identity operation)
Then there are two different states reaching the detector $A$, $|+1\rangle$ and $|+2\rangle$. Now the outcome probability can additionally depend on the choice.
This mechanism can explain difference observed in Vienna. However, it does not say anything about its physical origin (e.g. Pockels cell mechanics).
To quantify it one should rerun the experiment, examining the influence of Pockels cell. It will not only help to find a better quantum model
for violation of local realism but also ensure that the process is well described within known microscopic theory of optical setups.
Otherwise, this analysis will help to identify novel quantum effect which might need explanation beyond current theory.

In Munich \cite{munch}, the data also show discrepancy with the simplest Bell model. In this case, all Bell correlators should be equal for choice-independent efficiency and visibility while the data show clear inequality (about 4 standard deviations).
For instance $N(AB_{23}=+/-)=251/1012$ while $N(AB_{24}=+/-)=932/242$.
 In this case however, the explanation does not even need larger space. It may be just due to the fact the state in not exactly Bell's (maximally entangled). In general it can be anything in the $|\pm\pm\rangle$ basis. Moreover, the angles $\phi$ may deviate from the expected values.
Although this is a reasonable explanation, there is no comment and estimate about the discrepancy in the paper.

\section{Objectivity loophole}

Recent experiments made considerable progress in violation local realism because they closed simultaneously two important loopholes - detection and communication. Detection loophole was common for photon-based experiments with low detection efficiency. For small rate of detected photons the additional outcome $0$ (no-detection or lost photon) makes impossible to violate (\ref{chshin}). Violation is then only possible making an addition, unverifiable assumption of fair sampling, namely, that $0$ outcome occurs randomly and so it can be removed form analysis. On the other hand communication of locality loophole occurs whenever the distance between $A$ and $B$ is so small that the signal about the choice can reach the other party before completing the readout.

The recent experiments \cite{hensen,nist,vien, munch} cannot close the freedom of choice loophole, which means that the choice is in fact predetermined or simply determined earlier than claimed. An attempt to close it by cosmic rays \cite{cosm} opens again detection loophole because of low fraction of detected photons.
The loophole occurs also due to assumed readout time, determined experimentally by the time taggers.
The loophole is somewhat philosophical because there is no objective criterion for choice and readout times. Even cosmic photons will not be fully convincing,
after all one has to trust that the photons are indeed uncorrelated (and that they were not swapped with other, correlated ones).
Therefore, it is rather the question of reaching practical thresholds.

\begin{figure}
\includegraphics[scale=.5]{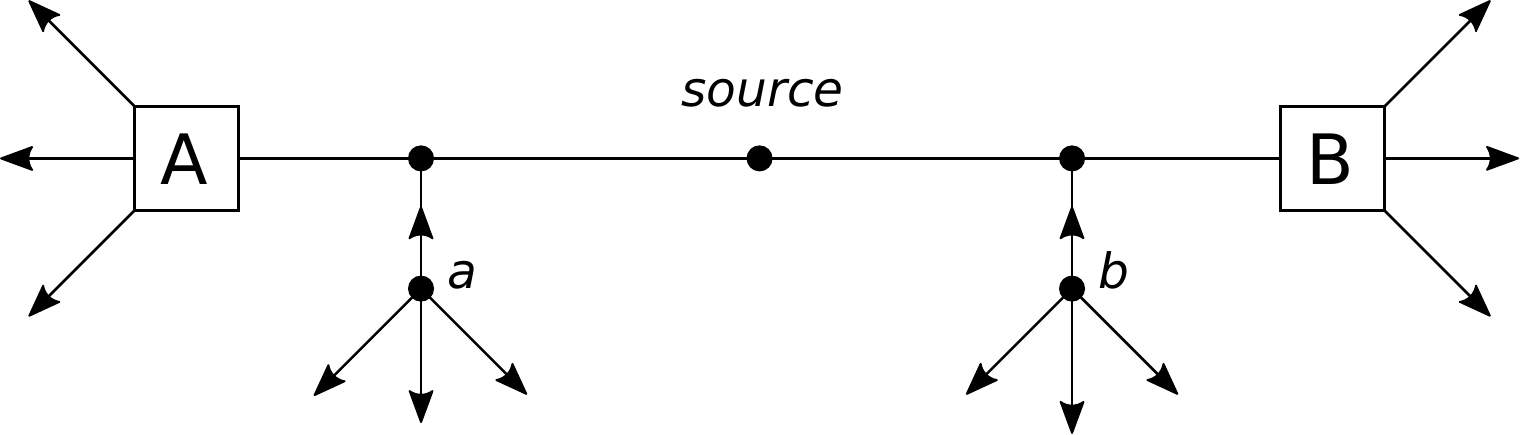}
\caption{Closing objectivity loophole in practice. Notation as in Fig. \ref{spt}.
 Information about both the choice and readout, including timetags, must be stored in several copies, preferably independent computers, indicated by arrows, except arrows indicating influence of the choice on the measurement (not vice versa!). Data processing 
after finishing the experimental run should include comparison between copies. All events with difference between copies must be excluded from considerations.}
\label{objl}
\end{figure}

However, one should obey the criterion of objective readout, neglected so far experimentally, i.e. the clonability of the outcome. Objective realism means that
we can copy the result of the measurements an arbitrary number of times (broadcasting) \cite{broad}. 
Operationally, it means that the positive operator-valued measure (POVM) \cite{povm} describing the measurement by creating a new outcome-dependent state 
 $\hat{\rho}\to\hat{K}\hat\rho\hat{K}^\dag$ (Kraus operators $\hat{K}$ \cite{kraus}), is not just a set of projections, $\hat{K}=\hat{P}_x=|x\rangle\langle x|$ with $\sum_x\hat{P}_x=\hat{I}$ (identity) in some basis $x$. Broadcasting or cloning  essentially demands a large set of copies $|x_1,x_2,...,x_m\rangle$ with the number of copies $m\to\infty$ and
\begin{equation}
\hat{K}_x=|x_1=x,x_2=x,...x_m=x\rangle\langle x|.
\end{equation}
 Without demonstrating clonability,  an experiment has the objectivity loophole left open. This means that a hacker, or simply some unspecified subluminal interaction
can change the registered value after the readout time tag using the information about the other party's choice. Then the inequality (\ref{chshin}) or (\ref{eber}) is violated but the communication has lasted longer and the final (hacker's) time tag is already within lightcone. Violation of Bell-type inequalities in this case does not refute local realism. Of course in experimental practice we can achieve only a finite number of copies. A fair approach to objectivity would require making at least two independent copies. 
It should be relatively simple to implement rerunning recent experiments. Once the readout is complete, it should be copied to an independent storage (computer)
with the final time tag after storing both copies, see Fig. \ref{objl}. In the data analysis one should check if the copies are the same. 
It will not only help to demonstrate closing the loophole but also prove security of the data. If violation of local realism is useful for quantum cryptography, the whole process must be protected from hacking, e.g. modifying the data after the readout. It may be less likely to 
hack two independent storages than a single one. 

\section{Conclusion}

Recent experimental Bell-type test rely on several, hard to verify assumptions, including relativity, freedom of choice, and objectivity, constituting formal loopholes.
We suggest to get free from relativity assumption by closing communication channels, and show objectivity by cloning the results immediately after each event.
At least the latter should not be technically demanding, requiring simply storing data on several computers. In addition, anomalies in the data need explanation by some diagnostic run, capturing yet unspecified effect of Pockels cells or other devices in the setup.

\section*{Acknowledgments}
The objectivity loophole has been inspired by discussion with J. Korbicz, P. Mironowicz and R. Horodecki in Sopot, Poland, in May 2017.

\end{document}